\documentclass[reprint, amsmath, amssymb, aps, superscriptaddress, prx]{revtex4-2}

\usepackage{bm,color,amsmath,txfonts}

\usepackage{graphicx}
\usepackage{subfigure}
\usepackage{verbatim}
\usepackage{dcolumn}
\usepackage{bm}
\usepackage{epsf}
\usepackage{xcolor}
\usepackage{hyperref}
\usepackage{hhline}
\usepackage{float}
\usepackage{enumerate}
\usepackage{bbm}
\usepackage{lipsum}
\usepackage{natbib}
\usepackage{soul}
\definecolor{mygreen}{rgb}{0.01, 0.31, 0.59}
\definecolor{myblue}{rgb}{0.01, 0.31, 0.59}
\definecolor{myred}{rgb}{0.63, 0.12, 0.12}
\hypersetup{
	colorlinks=true,   
	linkcolor=blue,  
	citecolor=blue, 
	urlcolor =blue    
}

\hypersetup{
	colorlinks=true,   
	linkcolor=blue,  
	citecolor=blue, 
	urlcolor =blue    
}

\usepackage{ulem}

\begin{document}
	
\title{Squeezed light in a semiconductor microcavity}

\author{Xuan Zuo}
\affiliation{Zhejiang Key Laboratory of Micro-Nano Quantum Chips and Quantum Control, School of Physics, and State Key Laboratory for Extreme Photonics and Instrumentation, Zhejiang University, Hangzhou 310027, China}
\author{Zi-Xu Lu}
\affiliation{Zhejiang Key Laboratory of Micro-Nano Quantum Chips and Quantum Control, School of Physics, and State Key Laboratory for Extreme Photonics and Instrumentation, Zhejiang University, Hangzhou 310027, China}
\author{Zhi-Yuan Fan}
\affiliation{Zhejiang Key Laboratory of Micro-Nano Quantum Chips and Quantum Control, School of Physics, and State Key Laboratory for Extreme Photonics and Instrumentation, Zhejiang University, Hangzhou 310027, China}
\author{Shi-Yao Zhu}
\affiliation{Zhejiang Key Laboratory of Micro-Nano Quantum Chips and Quantum Control, School of Physics, and State Key Laboratory for Extreme Photonics and Instrumentation, Zhejiang University, Hangzhou 310027, China}
\affiliation{College of Optical Science and Engineering, Zhejiang University, Hangzhou 310027, China}
\author{Jie Li}\thanks{Corresponding author: jieli007@zju.edu.cn}
\affiliation{Zhejiang Key Laboratory of Micro-Nano Quantum Chips and Quantum Control, School of Physics, and State Key Laboratory for Extreme Photonics and Instrumentation, Zhejiang University, Hangzhou 310027, China}

\begin{abstract}
\noindent\textbf{Abstract}\\
\noindent
Squeezed light is a particularly useful quantum resource, which finds broad applications in quantum information processing, quantum metrology and sensing, and biological measurements. Here we show how to produce squeezed light exploiting the strong exciton-phonon nonlinear interaction in a semiconductor microcavity. The semiconductor microcavity is embedded with a quantum well, which supports both linear and nonlinear interactions among excitons, phonons, and cavity photons. We show that the strong exciton-phonon deformation potential interaction can induce a quadrature-squeezed cavity output field, and further reveal an important role of the exciton-photon coupling in engineering the squeezing spectrum and improving the robustness of the squeezing against thermal noise. Our results indicate that substantial optical squeezing in a broad band, up to tens of gigahertz, can be achieved using currently available parameters.
\end{abstract}

\maketitle

\vspace{0.6cm}
\noindent\textbf{Introduction}\\
\noindent
Squeezed states of light are a type of optical nonclassical states that have less uncertainty in one quadrature than a coherent state~\cite{Walls}.  The generation of squeezed light typically requires certain optical nonlinear interaction~\cite{Leuchs}. It was first generated  by exploiting the process of four-wave-mixing in an atomic vapor of sodium atoms~\cite{Slusher}. Subsequently, many groups successfully demonstrated that squeezed light can also be generated in various optical systems, such as an optical fiber~\cite{fiber}, a nonlinear crystal~\cite{crystal}, a semiconductor laser~\cite{Itaya}, a single atom in an optical cavity~\cite{Rempe}, a single semiconductor quantum dot~\cite{Schulte}, an optomechanical system~\cite{Brooks12,Safavi-Naeini13,Purdy13}, etc.  Squeezed light is a vital quantum resource and finds very broad applications in quantum information science, and quantum metrology and sensing. For example, it can be used to improve the sensitivity in the gravitational wave detection~\cite{LIGO,RSch}, {qubit readout~\cite{Qin22,Qin24}}, and biological measurements~\cite{Taylor,GSA}, and to produce Einstein-Podolsky-Rosen entangled states for realizing various quantum protocols, such as quantum teleportation~\cite{Bouw,Furu}.

Here, we provide a new approach for producing squeezed light in a semiconductor microcavity system. Specifically, we consider a semiconductor microcavity sandwiched between distributed Bragg reflectors (DBRs), which can simultaneously confine light and sound waves. The microcavity is also embedded with a quantum well (QW), forming an exciton-phonon-photon tripartite system. One of the distinct advantages of this hybrid system is that it exhibits rich nonlinear interactions with adjustable strength, which can be exploited to generate novel quantum states, {including ground state cooling of mechanical motion~\cite{Bloch22} and excitonic entanglement~\cite{zuoprr,zuopra}}. In this work, we show how the nonlinear exciton-phonon (EPn) interaction can be used to reduce the quantum noise of light, and how the exciton-photon (EPt) coupling can provide a new degree of freedom to engineer the squeezing spectrum of the cavity output field and improve the robustness of the squeezing against environmental temperature. 
{Impressively, a substantial squeezing can be achieved using a very bad microcavity, where the strong-coupling induced exciton polaritons~\cite{Hopfield58} are not present.}

\vspace{0.6cm}
\noindent\textbf{Results}\\
\noindent\textbf{The system}\\
\noindent
We consider a semiconductor microcavity formed by DBRs, capable of confining both light and sound waves~\cite{Perrin13} (Fig.~\ref{fig1}(a)). The microcavity is embedded with a QW, and thus the system becomes an exciton-phonon-photon hybrid system~\cite{Kyriienko14,Sesin23}, which can be, e.g., a planar GaAs/AlAs microcavity grown by molecular beam epitaxy~\cite{Rozas14,Lemaitre15}.  In this tripartite system, any two parties can interact, either linearly or nonlinearly, and their coupling can be strong~\cite{Santos23}. 
Specifically, the electron and hole in an exciton form a dipole that can couple with microcavity photons, and the phonon-induced geometric deformation can directly change the cavity resonance frequency, or modify it via altering the refractive index of the material (i.e., the photoelastic effect), establishing an optomechanical coupling~\cite{Perrin13,Favero14}. Besides, the mechanical strain can also couple to excitons via the deformation potential (DP) interaction, which modifies the semiconductor band structure causing an excitonic frequency shift~\cite{Bir,Bloch22}.  
Recent studies indicate that the vacuum EPn coupling strength can be up to megahertz~\cite{Bloch22,Sesin23}. 
 It is worth emphasizing that the EPn (EPt) coupling strength depends on the position of the QW in the strain (cavity) field. Usually, the intensity of the cavity and strain fields do not overlap~\cite{Perrin13,Sesin23}, implying that the maximum EPn and EPt coupling cannot be achieved simultaneously. These two couplings can be adjusted by placing the QW at different positions.  In the studies of exciton polaritons~\cite{Hopfield58}, the QW is placed at the maximum cavity field to increase the EPt coupling strength~\cite{Weisbuch92}. {In this regime, an interesting mechanical self-oscillation was observed by exploiting the polariton-mechanical quadratic coupling~\cite{AF20}, which may lead to the squeezing of mechanical motion~\cite{AF22}}. Besides, the overlap between the optical intensity and the square displacement determines the photon-phonon coupling strength~\cite{Perrin13}. Therefore, different choices of the configuration of the QW and microcavity layers provide an effective means to regulate the coupling strength among the three modes. The great adjustability of multiple couplings is a notable advantage of this hybrid system.

 
The general Hamiltonian of the system, with a laser driving the microcavity,  is given by
\begin{align}\label{HHH}
	\begin{split}
		H/\hbar {=} \! & \sum_{j=a,b,d}  \! \omega_{j} j^\dagger j  \! + g_{\rm ad} \left(a^\dagger d+ad^\dagger \right) {-} g_{\rm ab} a^\dagger a \left( b + b^\dagger \right) \\
		&  \,\,\,  - g_{\rm db} d^\dagger d \left( b + b^\dagger \right) \! + i\Omega \left(a^\dagger e^{-i\omega_0t}-a e^{i\omega_0t} \right)  ,
	\end{split}
\end{align}
where $j=a, d, b$ denote the bosonic annihilation operators of the cavity, exciton, and phonon modes, respectively, satisfying the canonical commutation relation $[j, j^\dagger ] {=} 1$, and $ \omega_{j}$ are their corresponding resonance frequencies; $g_{\rm ad}$ is the EPt coupling strength, which can exceed both the cavity and exciton dissipation rates $\kappa_a$ and $\kappa_d$, leading to exciton polaritons~\cite{Weisbuch92}; $g_{\rm ab}$ denotes the bare optomechanical coupling strength due to the radiation-pressure or photoelastic interaction~\cite{Perrin13,Favero14}; $g_{\rm db}$ represents the bare EPn coupling strength arising from the DP interaction~\cite{Bir,Bloch22}; and the last term describes the microcavity driven by a monochromatic laser field. The coupling strength between the drive field and the cavity is {$\Omega = \sqrt{P \kappa_1/ \hbar \omega_0}$}, where $P$ ($\omega_0$) is the power (frequency) of the laser, and $\kappa_1$ is the cavity external decay rate, whereas $\kappa_2 \equiv \kappa_a - \kappa_1$ denotes the decay rate accounting for all other dissipation channels, such as absorption and scattering losses within the cavity.

We adopt the quantum Langevin equations (QLEs) to account for the coupling of the system to the external environment, which include the dissipation and input noise of each mode. In the frame rotating at the laser frequency, we obtain
\begin{align}\label{QLEAA}
	\begin{split}
			\dot{a} \! =&\! - \left(\frac{\kappa_a}{2} \! + i{\Delta}_a \right) a \!- i g_{\rm ad} d \!+ i g_{\rm ab} a \left(b \!+ b^\dagger \right) \!+ \Omega \!+ \sum_{l=1,2} \sqrt[]{\kappa_l} a_l^{in} ,  \\
			\dot{d} \! =&\! - \left(\frac{\kappa_d}{2} \! + i{\Delta}_d \right) d \!- i g_{\rm ad} a \!+ i g_{\rm db} d \left(b \!+ b^\dagger \right) + \sqrt[]{\kappa_d} d^{in},  \\
			\dot{b} \! =&\! - \left(\frac{\kappa_b}{2} + i{\omega}_b \right) b \! + i g_{\rm ab} a^\dagger a + i g_{\rm db} d^\dagger d + \sqrt[]{\kappa_b} b^{in},
	\end{split}
\end{align}
where ${\Delta}_{a(d)} = \omega_{a(d)} - \omega_0$, $\kappa_b$ is the mechanical damping rate, and $\mathcal{R}^{in}$ $(\mathcal{R}=a_{l},d,b)$ denote the input noise operators, which are zero-mean and characterized by the correlation functions~\cite{Zoller}: $\langle \mathcal{R}^{in}(t) \mathcal{R}^{in\dagger}(t^\prime) \rangle=(N_j+1)\delta(t-t^\prime)$, $\langle \mathcal{R}^{in\dagger}(t) \mathcal{R}^{in}(t^\prime) \rangle=N_j\delta(t-t^\prime)$, with $N_j=\big[\! \exp[(\hbar \omega_j/k_BT)]{-}1\big]^{-1}$ being the equilibrium mean thermal excitation number of the mode $j$, $k_B$ being the Boltzmann constant, and $T$ as the bath temperature.

A strong laser driving field gives rise to large amplitudes of the cavity and exciton modes in the steady state, i.e., $|\langle a \rangle|, |\langle d \rangle| \gg 1$, where the steady-state averages are
\begin{align}\label{AveSt}
	\begin{split}
\langle a \rangle &= \frac{\left( \frac{\kappa_d}{2} + i {\tilde{\Delta}}_d \right) \Omega}{g_{\rm ad}^2 + \left( \frac{\kappa_a}{2} + i \tilde{\Delta}_a \right) \left( \frac{\kappa_d}{2} + i {\tilde{\Delta}}_d \right)}, \\ 
\langle d \rangle &= \frac{- i g_{\rm ad} \langle a \rangle}{ \frac{\kappa_d}{2} + i {\tilde{\Delta}}_d }, \,\,
\langle b \rangle = \frac{i g_{\rm ab} | \langle a \rangle|^2 + i g_{\rm db} | \langle d \rangle|^2}{  \frac{\kappa_b}{2} + i \omega_b }, \\
\end{split}
\end{align}
with $\tilde{\Delta}_d = {\Delta}_d + 2 g_{\rm db} {\rm Re}[\langle b \rangle]$ ($\tilde{\Delta}_a = {\Delta}_a + 2 g_{\rm ab} {\rm Re}[\langle b \rangle]$) being the effective exciton (cavity)-drive detuning by including the frequency shift induced by the DP (optomechanical) interaction. This allows us to linearize the dynamics of the quantum fluctuations $\delta j =j- \langle j \rangle$ ($j=a, d, b$) around the large steady-state averages by neglecting small second-order fluctuation terms. The linearized QLEs for the quantum fluctuations are
\begin{align}\label{QLEAAfluc}
	\begin{split}
		\delta \dot{a}  =& - \left(\frac{\kappa_a}{2} + i{\tilde{\Delta}}_a \right) \delta a  - i g_{\rm ad} \delta d \\+& i G_{\rm ab} \left(\delta b + \delta b^\dagger \right) + \sum_{l=1,2} \sqrt[]{\kappa_l} a_l^{in} ,  \\
		\delta \dot{d}  =& - \left(\frac{\kappa_d}{2} + i{\tilde{\Delta}}_d \right) \delta d  - i g_{\rm ad} \delta a \\+& i G_{\rm db} \left(\delta b + \delta b^\dagger \right) + \sqrt[]{\kappa_d} d^{in},  \\
		\delta \dot{b}  =& - \left(\frac{\kappa_b}{2} + i{\omega}_b \right) \delta b + i \left(G_{\rm ab}^* \delta a + {\rm H.c.} \right) \\ + & i \left(G_{\rm db}^* \delta d + {\rm H.c.} \right) + \sqrt[]{\kappa_b} b^{in},
	\end{split}
\end{align}
where $G_{\rm ab} = g_{\rm ab} \langle a \rangle$ and $G_{\rm db} = g_{\rm db} \langle d \rangle$ are the effective photon-phonon and EPn coupling strength, which are significantly enhanced due to a large number of cavity photons and excitons.

The QLEs~\eqref{QLEAAfluc} can be conveniently solved in the frequency domain by taking the Fourier transform of each equation using
\begin{align}
\begin{split}
		\delta j(\omega&) = \int^{+\infty}_{-\infty} \!\! \delta j(t) e^{i \omega t} {\rm d}t, \\
		\delta j^\dagger(-\omega&) \,\,{= [\delta j(-\omega)]^\dagger} = \int^{+\infty}_{-\infty} \!\! \delta j^\dagger(t) e^{i \omega t} {\rm d}t,
	\end{split}
\end{align}
and $- i \omega \delta j(\omega) = \int^{+\infty}_{-\infty} \delta \dot{j}(t) e^{i \omega t} {\rm d}t$.  The solution of the cavity field fluctuation $\delta a(\omega)$ can then be achieved, which is a function of the input noise terms, but is too lengthy to be reported here.

{Our aim is to generate stationary squeezing of the output optical field, so the system stability must be analyzed. A detailed analysis is provided in the supplementary material.}
 The fluctuation of the cavity output field can be obtained using the input-output relation, $\delta a^{\rm out}(\omega) = \sqrt{\kappa_1} \delta a(\omega) - a_1^{\rm in}(\omega)$, and a general quadrature of the output field is defined as
\begin{align}
	\begin{split}
		\delta X^{\rm out}_{\phi} (\omega) \!= \frac{1}{\sqrt{2}}\left[ \delta a^{\rm out} (\omega) e^{- i \phi} \!+ \delta a^{{\rm out} \dagger} (-\omega) e^{i \phi}\right] ,
	\end{split}
\end{align}
with $\phi$ being the phase angle, which is related to the phase of the local oscillator in the Homodyne detection. The noise spectral density (NSD) of the output field quadrature is given by
\begin{align}
	\begin{split}
		S^{\rm out}_{\phi} (\omega) \!= \frac{1}{2 \pi} \int^{+\infty}_{-\infty} d \omega^{\prime} e^{- i \left(\omega + \omega^{\prime}\right) t}\left< \delta X^{\rm out}_{\phi} (\omega) \delta X^{\rm out}_{\phi} (\omega^{\prime}) \right> ,
	\end{split}
\end{align}
{which requires the use of the input noise correlations in the frequency domain: $\langle \mathcal{R}^{in}(\omega) \mathcal{R}^{in\dagger}(-\omega^\prime) \rangle=2\pi(N_j+1)\delta(\omega+\omega^\prime)$, $\langle \mathcal{R}^{in\dagger}(-\omega) \mathcal{R}^{in}(\omega^\prime) \rangle=2 \pi N_j\delta(\omega+\omega^\prime)$. This leads to $\omega^\prime = -\omega$ during the integration, thereby rendering the NSD time-independent.}
In our definition, the vacuum fluctuation corresponds to $S_{\rm vac} (\omega)= 0.5$, and thus $S^{\rm out}_{\phi} (\omega) <0.5$ indicates that the fluctuation of the output field quadrature at the phase angle $\phi$ is below the vacuum level, i.e., the output field is squeezed. The degree of squeezing is quantified in the dB unit via $S = -10 \log_{10} [S^{\rm out}_{\phi} (\omega)/S_{\rm vac}(\omega) ]$.

\vspace{0.3cm}
\noindent\textbf{Squeezing of quantum noise}\\
\noindent
A recent experiment has achieved a remarkably strong EPn bare coupling $g_{\rm db}/2\pi \simeq 20$~MHz~\cite{Sesin23}, implying that a strong effective coupling $G_{\rm db}$ can be achieved by increasing the drive power (before the system enters the unstable regime~\cite{DeJesus87}). The potentially strong EPn nonlinear interaction is a salient feature of this hybrid system, which we show below can be used to reduce the quantum noise of light.  To distinguish it from the optomechanical ponderomotive squeezing~\cite{Fabre94,Tombesi94,Brooks12,Safavi-Naeini13,Purdy13}, we consider a configuration in which the optomechanical coupling is designed to be negligibly small, i.e., $|G_{\rm ab}| \ll |G_{\rm db}|$. We first show the results of squeezing caused solely by the EPn interaction (via setting $G_{\rm ab}=0$), and then analyze the effect of the residual optomechanical coupling ($G_{\rm ab}\neq0$)  on the squeezing, due to the non-perfect elimination of the coupling in the experiment.

The EPn interaction describes the coupling between the mechanical displacement and the exciton number. Such an intensity-dependent displacement will cause a frequency shift (thus a phase shift) of the exciton mode. Consequently, the amplitude and phase quadratures of the exciton mode get correlated via the mediation of phonons, leading to the quadrature squeezing of the exciton mode. This excitonic squeezing, being experimentally inaccessible, can however transfer to the intracavity field through the EPt beamsplitter (state-swap) interaction, leading to a squeezed cavity output field, which can be verified by measuring the NSD of the output field. We note that the EPt beamsplitter coupling itself does not generate squeezing, so the squeezing in the output field can only result from the nonlinear EPn interaction.  

{
In Fig.~\ref{fig2}(a), we plot the degree of squeezing of the output field quadrature, which exhibits a pronounced squeezing around the mechanical frequency $\omega \simeq \omega_b$, with a maximum of $\sim$7 dB below the vacuum fluctuation at an optimal $\phi$. 
In getting Fig.~\ref{fig2}(a), we used a slightly red-detuned laser field with respect to the cavity and exciton modes, $0<\tilde{\Delta}_a, \tilde{\Delta}_d \ll \omega_b$, which helps to stabilize the system by realizing an effective cooling of the phonon mode, corresponding to the EPn anti-Stokes scattering outperforming the Stokes one. Nevertheless, a much larger detuning close to the mechanical frequency, though optimal for realizing mechanical cooling, reduces the degree of squeezing, as shown in Figs.~\ref{fig2}(b) and~\ref{fig2}(c). A too small exciton-drive detuning, e.g., $\tilde{\Delta}_d < 0.05 \omega_b$, should also be avoided as it increases the strength of the Stokes scattering, which makes the system tend to be unstable, as shown in Fig.~\ref{fig2}(c)}. 
The cavity and exciton modes should be close in frequency: As analyzed above, the output squeezing is transferred from the exciton mode, and the optimal state transfer requires the two modes to be nearly resonant~\cite{Li19,Yu20}.  In getting Fig.~\ref{fig2}, we have employed the following experimentally accessible parameters~\cite{Perrin13,Bloch22,Sesin23}: $\omega_b/2\pi = 20$ GHz, $\omega_d/2\pi = 360$ THz, $\kappa_a/2\pi = 20$ GHz ($\kappa_1 =9\kappa_2= 0.9\kappa_a$), $\kappa_b/2\pi = 1$ MHz, $\kappa_d/2\pi = 2$ GHz, $g_{\rm ad}/2\pi = 20$ GHz, $g_{\rm db}/2\pi = 20$ MHz, and $T = 4$ K. We take $|G_{\rm db}|/2\pi = 4$~GHz, which corresponds to $\Omega/2\pi=4$ THz and a laser power {$P \simeq 1.3$~mW}.  {It is worth noting that  within the stable region in Fig.~\ref{fig2}, the effective phonon number is below 4, indicating that the fluctuations of the phonon mode also remain small. }

The desired squeezed light requires a high degree of squeezing in a broad bandwidth. The degree of squeezing can be improved by increasing the EPn cooperativity ${\cal C} = \frac{4 |G_{\rm db}|^2}{\kappa_d \kappa_b}$ (Fig.~\ref{fig3}(a)). Although Fig.~\ref{fig2} shows a strong squeezing $\sim7$ dB, the bandwidth is relatively narrow for the degree of squeezing larger than, e.g., 3 dB. Nevertheless, the bandwidth can be greatly increased by optimizing the EPt coupling strength $g_{\rm ad}$ to yield a significantly improved EPn cooperativity ${\cal C}$. Equation~\eqref{AveSt} indicates that the steady-state exciton number $|\langle d \rangle|^2$ and thus the cooperativity ${\cal C} \equiv \frac{4 g_{\rm db}^2 }{\kappa_d \kappa_b}|\langle d \rangle|^2$ are determined by the coupling $g_{\rm ad}$, and their dependencies are shown in Fig.~\ref{fig3}(b). Clearly, for a given drive power there is an optimal EPt coupling strength corresponding to a maximal exciton number and EPn cooperativity. Compared to the squeezing spectrum of Fig.~\ref{fig2} with a large but non-optimal coupling $g_{\rm ad}/2\pi = 20$~GHz (i.e., the blue curve of Fig.~\ref{fig3}(c)), the significantly improved cooperativity at an optimal coupling $g_{\rm ad}/2\pi \simeq 8$~GHz yields an increased bandwidth, {about one mechanical frequency}, within which the squeezing remains strong (the red curve of Fig.~\ref{fig3}(c)). {This is much wider than the bandwidth, at the level of $\sim$ 10 MHz, of the squeezed light generated using a nonlinear crystal in an optical cavity (about 15 dB)~\cite{Vahlbruch16} or exploiting four-wave mixing in an atomic ensemble (about 10 dB)~\cite{Liu19}.} {In this situation, the exciton-photon coupling strength is smaller than the cavity decay rate ($g_{\rm ad} < \kappa_a$), and thus exciton polaritons do not form, rendering the polaromechanical platform~\cite{Santos23} not optimal for our protocol for squeezed light.} Differently, the maximum squeezing is centered at the zero frequency $\omega=0$, rather than at the mechanical frequency $\omega \simeq \pm \omega_b$ as in the ponderomotive squeezing~\cite{Fabre94,Tombesi94,Brooks12,Safavi-Naeini13,Purdy13}. Fig.~\ref{fig3}(c) indicates that the EPt coupling plays an essential role in shaping the profile of the squeezing spectrum, and therefore offers a new degree of freedom to engineer the squeezing spectrum on demand, which is a major advantage of our tripartite system over the bipartite optomechanical system.


In Figs.~\ref{fig4}(a) and~\ref{fig4}(b), we investigate the impact of the exciton and cavity dissipation rates on the degree of squeezing. In general, the smaller the exciton dissipation rate, the better the squeezing performance, because this increases the EPn cooperativity and hence the degree of squeezing (Fig.~\ref{fig3}(a)). For the state-of-art of the exciton decay rate $\kappa_d/2\pi\sim10^2$ MHz~\cite{Deng10,Sesin23}, a substantial squeezing close to 10~dB is potentially achievable (Fig.~\ref{fig4}(a)).  There are, however, two effects related to the cavity decay rate $\kappa_a$: On the one hand, $\kappa_a$ should be small as a larger $\kappa_a$ induces a larger effective exciton decay rate (due to the cavity-exciton coupling) and thus reduces the EPn cooperativity; on the other hand, $\kappa_a$, essentially $\kappa_1$, should be sufficiently large as it determines the pump efficiency and the output of the cavity field. The trade-off between these two effects leads to an optimal $\kappa_a$ for the squeezing, as shown in Fig.~\ref{fig4}(b).

We now analyze the effect of the residual optomechanical coupling $G_{\rm ab}$ on the EPn-induced squeezing. In practice, semiconductor microcavities inevitably exhibit some degree of the optomechanical coupling.  Nevertheless, the optomechanical coupling can be designed to be negligibly small with a bare coupling $g_{\rm ab} \ll g_{\rm db}$~\cite{Sesin23}.  Fig.~\ref{fig4}(c) shows the results by including the optomechanical interaction in the model ($g_{\rm ab}\neq0$).  The squeezing is slightly reduced due to an interference effect between the EPn and optomechanical interactions, which are both of the dispersive type. 
Nonetheless, our results indicate that significant squeezing can still be achieved even in the presence of some moderate optomechanical coupling. 
Fig.~\ref{fig4}(d) manifests the robustness of the squeezing against the bath temperature, and moderate squeezing could be achieved even at room temperature $T = 300$ K.
{At this high temperature, the phonon mode can still be cooled close to its ground state with an effective phonon number about 4.3, due to a strong EPn cooperativity.}
Moreover, it reveals another superiority of the EPt coupling: By optimizing $g_{\rm ad}$, not only the squeezing bandwidth is greatly increased, but also the robustness of the squeezing to thermal noise is significantly improved, as is shown in the inset of Fig.~\ref{fig4}(d) that the squeezing only experiences a slight reduction when the temperature rises by two orders of magnitude.  This could be understood that the maximum squeezing closer to the mechanical sideband is more affected by thermal noise as the phonon mode has the lowest frequency in the system and thus most thermal noise. 
We remark that although high temperatures may overcome the binding energy of excitons and destabilize them (e.g., in the GaAs materials with the binding energy of $\sim$10 meV, excitons will get ionized for the temperature higher than $\sim$116 K), room-temperature squeezing of light remains possible if using a material with high exciton binding energy~\cite{Lidzey98,Sun08,Kena-Cohen08,Su21,IG25}.

\vspace{0.3cm}
\noindent\textbf{Discussion}\\
\noindent
Since the exciton-exciton nonlinearity may be present in the system, such as the self-Kerr type with the Hamiltonian $H_{\rm Kerr} = \hbar K d^\dag d^\dag d d$~\cite{Bloch22} ($K$ is the Kerr coefficient), we now discuss its impact on the squeezed light in our protocol. As is known, the self-Kerr nonlinearity results in a frequency shift of the exciton mode, $\delta \omega_d = 2K |\langle d \rangle|^2$, and this further shifts the frequency of the optimal squeezing in the spectrum, as shown in Fig.~\ref{fig5}. Clearly, stronger Kerr effect leads to a larger frequency shift, but it does not appreciably reduce the degree of squeezing.  
The details of the calculation including the exciton self-Kerr nonlinearity are provided in the supplementary material.

We have presented a mechanism to generate squeezed light based on the nonlinear exciton-phonon deformation potential interaction in a semiconductor microcavity. Significant squeezing of the cavity output field can be achieved for a sufficiently large exciton-phonon cooperativity. The exciton-photon coupling provides an effective means to engineer the squeezing spectrum, and can greatly increase the squeezing bandwidth and improve the robustness of the squeezing to thermal noise. The protocol is within reach of current technology and may become a promising new approach for producing {broadband squeezed light, serving as a critical resource in quantum information applications; for example, it enables high clock frequencies in continuous-variable quantum computing~\cite{Yokoyama13,Kashiwazaki23} and enhances the channel capacity and communication rate in continuous-variable quantum communication~\cite{Shi20,Liang24}.}

\vspace{0.3cm}
\noindent
\textbf{Methods}\\
\noindent\textbf{Analysis of system stability and effective mean phonon number}\\
\noindent
Here we show in detail how the stability of the system is analyzed. This ensures that all the results presented in the main text are in the steady state.  The linearized QLEs for the quantum fluctuations, i.e., Equation (4) in the main text, after introducing the quadrature fluctuations $\delta X_O = (\delta O + \delta O^{\dagger})/\!\sqrt{2}$ and $\delta Y_O = i (\delta O^{\dagger} - \delta O)/\!\sqrt{2}$ ($O = a, d, b$),  can be cast in the following compact form:
\begin{align}
	\begin{split}\label{dotu}
	\dot{u}(t)={\cal R}u(t) + n(t),
	\end{split}
\end{align}
where $u(t)=\big[\delta X_a(t),\delta Y_a(t),\delta X_d(t),\delta Y_d(t),\delta X_b(t),\delta Y_b(t) \big]^{\rm T}$ is the vector of quadrature fluctuations, $n(t)=\big[\sqrt[]{\kappa_1}X_{a_1}^{in}+\sqrt[]{\kappa_2}X_{a_2}^{in},\sqrt[]{\kappa_1}Y_{a_1}^{in}+\sqrt[]{\kappa_2}Y_{a_2}^{in},\sqrt[]{\kappa_d}X_d^{in},\sqrt[]{\kappa_d}Y_d^{in},$ $\sqrt[]{\kappa_b}X_{b}^{in},\sqrt[]{\kappa_b}Y_{b}^{in} \big]^{\rm T}$ is the vector of input noises in the quadrature form, and the drift matrix ${\cal R}$ is given by
\begin{widetext}
\begin{align}
	\cal R=\begin{pmatrix}
		-\frac{\kappa_a}{2} & {\tilde{\Delta}}_a & 0 & g_{\rm ad} & -2 \mathrm{Im}[G_{\rm ab}] & 0 \\
		-{\tilde{\Delta}}_a & -\frac{\kappa_a}{2} & -g_{\rm ad} & 0 & 2 \mathrm{Re}[G_{\rm ab}] & 0\\
		0 & g_{\rm ad} & -\frac{\kappa_d}{2} & {\tilde{\Delta}}_d & -2 \mathrm{Im}[G_{\rm db}] & 0 \\
		-g_{\rm ad} & 0 & -{\tilde{\Delta}}_d & -\frac{\kappa_d}{2} & 2 \mathrm{Re}[G_{\rm db}] & 0\\
		0 & 0 & 0 & 0 & -\frac{\kappa_b}{2} & \omega_b\\
		2 \mathrm{Re}[G_{\rm ab}] & 2 \mathrm{Im}[G_{\rm ab}] & 2 \mathrm{Re}[G_{\rm db}] & 2 \mathrm{Im}[G_{\rm db}] & -\omega_b & -\frac{\kappa_b}{2}
	\end{pmatrix},
\end{align}
\end{widetext}
which governs the dynamics of the system.  The system becomes stable when all the eigenvalues of the drift matrix ${\cal R}$ possess negative real parts. This stability condition is guaranteed when we assign values to system parameters to obtain Figs. 2-4, that is, all the results are in the steady state.

We further show how to achieve the effective mean phonon number. This allows us to confirm that the phonon mode being cooled close to its ground state is a prerequisite to obtain strong squeezing of the cavity output field.  The steady state of the quadrature fluctuations is a three-mode Gaussian state, as a result of the linearized dynamics and the Gaussian nature of the input noises. Therefore, it can be fully characterized by a $6\times6$ covariance matrix (CM) $V$ with the entries defined as $V_{ij}=\frac{1}{2} \langle u_i(t)u_j({t}) + u_j({t})u_i(t) \rangle$ $(i,j=1,2,...,6)$. 
The corresponding steady-state CM $V$ can be achieved by solving the following Lyapunov equation~\cite{Vitali07}
\begin{align}
	\begin{split}
	{\cal R} V+V{\cal R}^T = -D,
	\end{split}
\end{align}
where $D$ is the diffusion matrix defined via $D_{ij}\,\delta(t \,{-} \,t')=\langle n_i(t)n_j(t')+n_j(t')n_i(t) \rangle/2$, and takes the specific diagonal form ${\rm diag}[\kappa_a \left(N_a {+} 1/2 \right), \kappa_a \left(N_a {+} 1/2 \right), \kappa_d \left(N_d {+} 1/2 \right), \kappa_d \left(N_d {+} 1/2 \right),\\ \kappa_b \left(N_b {+} 1/2 \right), \kappa_b \left(N_b {+} 1/2 \right)]$.

The energy of the phonon mode can then be calculated from the variances of the quadratures associated with the phonon mode, i.e., 
\begin{align}\label{UUU}
	\begin{split}
		U_b &= \hbar \omega_b \frac{\langle \delta X_{b}^2 \rangle + \langle \delta Y_{b}^2 \rangle}{2} \equiv \hbar \omega_b \left(N_{b,{\rm eff}} + \frac{1}{2} \right).
	\end{split}
\end{align}
Here, $N_{b,{\rm eff}}$ is the effective mean phonon number, which from Eq.~\eqref{UUU} is obtained as
\begin{align}
	\begin{split}
		N_{b,{\rm eff}} &= \frac{\langle \delta X_{b}^2 \rangle + \langle \delta Y_{b}^2 \rangle -1}{2}.
	\end{split}
\end{align}

\vspace{0.3cm}
\noindent\textbf{Effect of the exciton self-Kerr nonlinearity}\\
\noindent
In this appendix, we provide necessary details for the calculation incorporating the exciton self-Kerr type nonlinearity, which lead to the results of Figure 5 in the main text.  The Hamiltonian of the exciton-phonon cavity QED system is then given by
\begin{align}\label{HHH}
	\begin{split}
		H/\hbar =  & \sum_{j=a,b,d}  \omega_{j} j^\dagger j  + g_{\rm ad} \left(a^\dagger d+ad^\dagger \right)  - g_{\rm ab} a^\dagger a \left( b + b^\dagger \right) \\+& K d^\dagger d^\dagger d d - g_{\rm db} d^\dagger d \left( b + b^\dagger \right)   + i\Omega \left(a^\dagger e^{-i\omega_0t}-a e^{i\omega_0t} \right)  ,
	\end{split}
\end{align}
where $K$ is the exciton self-Kerr coefficient. The other parameters are introduced in the main text. The above Hamiltonian leads to the following quantum Langevin equations for the system, which in the frame rotating at the laser drive frequency are:
\begin{align}\label{QLEAA}
	\begin{split}
			\dot{a} \! =&\! - \left(\frac{\kappa_a}{2} \! + i{\Delta}_a \right) a - i g_{\rm ad} d + i g_{\rm ab} a \left(b + b^\dagger \right) + \Omega + \sum_{l=1,2} \sqrt[]{\kappa_l} a_l^{in} ,  \\
			\dot{d} \! =&\! - \left(\frac{\kappa_d}{2} \! + i{\Delta}_d \right) d - i g_{\rm ad} a + i g_{\rm db} d \left(b + b^\dagger \right) {-2 i K d^\dagger d d} + \sqrt[]{\kappa_d} d^{in},  \\
			\dot{b} \! =&\! - \left(\frac{\kappa_b}{2} + i{\omega}_b \right) b \! + i g_{\rm ab} a^\dagger a + i g_{\rm db} d^\dagger d + \sqrt[]{\kappa_b} b^{in}.
	\end{split}
\end{align}
Following the same linearization treatment used in the main text, we obtain the following solutions for the steady-state averages, given by
\begin{align}\label{AveSt}
	\begin{split}
\langle a \rangle &= \frac{\left( \frac{\kappa_d}{2} + i ({\tilde{\Delta}}_d {+ 2 K |\langle d \rangle|^2}) \right) \Omega}{g_{\rm ad}^2 + \left( \frac{\kappa_a}{2} + i \tilde{\Delta}_a \right) \left( \frac{\kappa_d}{2} + i ({\tilde{\Delta}}_d {+ 2 K |\langle d \rangle|^2}) \right)}, \\ 
\langle d \rangle &= \frac{- i g_{\rm ad} \langle a \rangle}{ \frac{\kappa_d}{2} + i ({\tilde{\Delta}}_d {+ 2 K |\langle d \rangle|^2}) }, \,\,\,\,\,
\langle b \rangle = \frac{i g_{\rm ab} | \langle a \rangle|^2 + i g_{\rm db} | \langle d \rangle|^2}{  \frac{\kappa_b}{2} + i \omega_b }.
\end{split}
\end{align}
By comparing with the solutions of Eq. 3 in the main text, it is clear that the self-Kerr nonlinearity gives rise to a frequency shift of the exciton mode, which is $\delta \omega_d = 2 K | \langle d \rangle|^2 $.  

We further obtain the linearized quantum Langevin equations for the quantum fluctuations of the three modes
\begin{align}\label{QLEAAfluc}
	\begin{split}
		\delta \dot{a} \! =&\! - \left(\frac{\kappa_a}{2} + i{\tilde{\Delta}}_a \right) \delta a \! - i g_{\rm ad} \delta d \!+ i G_{\rm ab} \left(\delta b + \delta b^\dagger \right) + \sum_{l=1,2} \sqrt[]{\kappa_l} a_l^{in} ,  \\
		\delta \dot{d} \! =&\! - \left(\frac{\kappa_d}{2} + i{\tilde{\Delta}}_d \right) \delta d \! - i g_{\rm ad} \delta a \!+ i G_{\rm db} \left(\delta b + \delta b^\dagger \right) \\-&4 i K |\langle d \rangle|^2 \delta d - 2 i K \langle d \rangle^2 \delta d^\dagger + \sqrt[]{\kappa_d} d^{in},  \\
		\delta \dot{b} \! =&\! - \left(\frac{\kappa_b}{2} + i{\omega}_b \right) \delta b \!+\!  i \left(G_{\rm ab}^* \delta a \!+\! {\rm H.c.} \right) \!+\! i \left(G_{\rm db}^* \delta d \!+\! {\rm H.c.} \right) \!+ \sqrt[]{\kappa_b} b^{in}.
	\end{split}
\end{align}
Solving the above equations in the frequency domain by taking the Fourier transform and utilizing the cavity input-output relation, we can obtain the fluctuations of the cavity output field. Following the same definitions as in the main text, we achieve its general quadrature $\delta X^{\rm out}_{\phi} (\omega)$ and the associated noise spectral density $S^{\rm out}_{\phi} (\omega)$.

\vspace{0.1cm}
\noindent
\textbf{Data availability}\\
\noindent All relevant data are within the paper. \\

\vspace{0.1cm}
\noindent
\textbf{Acknowledgments}\\
We thank Yaohui Zheng for useful discussion on squeezed light. This study was supported by National Key Research and Development Program of China (Grant No. 2022YFA1405200, 2024YFA1408900), National Natural Science Foundation of China (Grant No. 12474365, 92265202), and Zhejiang Provincial Natural Science Foundation of China (Grant No. LR25A050001). \\

\vspace{0.1cm}
\noindent
\textbf{Author contributions} \\
J.L. conceived the idea, X.Z. conducted the calculations, and X.Z., Z.X.L., Z.Y.F., S.Y.Z., and J.L. analyzed the results. X.Z. and J.L. wrote the manuscript and all authors commented on the manuscript. All authors have read and approved the manuscript.\\

\vspace{0.1cm}
\noindent
\textbf{Competing interests} \\
The authors declare no competing financial or non-financial interests.\\



\vspace{0.1cm}		
\noindent
\textbf{References}\\

\onecolumngrid
\clearpage
\begin{figure*}[h]
	\centering
	\caption{\textbf{The system and coupling structure.} (a) A semiconductor microcavity embedded with a QW supports an optical cavity mode, a phonon mode, and an exciton mode. (b) The phonon mode $b$ couples to the exciton mode $d$ via the DP interaction and to the optical cavity $a$ via the optomechanical interaction, and the exciton and cavity modes are coupled through the dipole interaction. The microcavity is driven by a monochromatic laser to enhance the exciton-phonon dispersive interaction, which induces the squeezing of the exciton mode. It further leads to the squeezing of the cavity field via the exciton-photon state swapping, and consequently a squeezed cavity output field. } 
	\label{fig1}
\end{figure*}

\begin{figure*}[h]
	\centering
	\caption{\textbf{Spectrum of the squeezed output field.} Degree of squeezing (dB) of the output field quadrature versus frequency $\omega$ for various (a) phase angles $\phi$ = $0.65$, $0.7$, $0.75$, $0.8$, and $0.85 \pi$; (b) effective cavity-drive detuning $\tilde{\Delta}_a$ = $-0.4$, $-0.2$, $0$, $0.2$, and $0.4 \omega_b$; (c) effective exciton-drive detuning $\tilde{\Delta}_d$ = $-0.1$, $0.1$, $0.3$, $0.5$, and $0.7 \omega_b$. The grey area in (c) corresponds to the unstable regime. We take $\tilde{\Delta}_d= 0.3\omega_b$ in (a) and (b), $\tilde{\Delta}_a= 0.1\omega_b$ in (a) and (c), and $\phi=0.75 \pi$ in (b) and (c). See text for the other parameters.}
	\label{fig2}
\end{figure*}

\begin{figure*}[h]
	\centering
	\caption{\textbf{Analysis of the cooperativity and coupling strength on the squeezing.} (a) Maximal degree of squeezing in the output spectrum versus exciton-phonon cooperativity $\cal C$. The cooperativity is changed by varying the exciton-phonon bare coupling $g_{\rm db}$ from 0 to 20 MHz for a fixed drive power.  (b) Steady-state intracavity photon number (dotted line) and exciton number (dashed line) and exciton-phonon cooperativity $\cal C$ (solid line) versus exciton-photon coupling $g_{\rm ad}$ under a fixed drive power. (c) Squeezing spectrum (in dB) of the output field quadrature for different values of the exciton-photon coupling: $g_{\rm ad}/2\pi = 20$, 15, 10, 8, and 6 GHz. The phase angle $\phi$ is optimized for squeezing for each cooperativity in (a) and each exciton-photon coupling in (c). We take $\tilde{\Delta}_a= 0.1\omega_b$, $\tilde{\Delta}_d= 0.3\omega_b$, and the other parameters are the same as in Fig.~\ref{fig2}(a).}
	\label{fig3}
\end{figure*}

\begin{figure*}[h]
	\centering
	\caption{\textbf{Effects of system dissipations and the optomechanical coupling.} Squeezing spectrum (in dB) of the output field quadrature for different values of (a) the exciton dissipation rate $\kappa_d$; (b) the external cavity decay rate $\kappa_1\equiv \kappa_a-\kappa_2$ ($\kappa_2/2\pi=2$ GHz); (c) the effective optomechanical coupling strength $|G_{\rm ab}|$ via changing $g_{\rm ab}$: $|G_{\rm ab}|/2\pi = 0$, 120, 240 MHz correspond to $g_{\rm ab} =0$, $0.1g_{\rm db}$, $0.2g_{\rm db}$, respectively; (d) the bath temperature $T$. In (a)-(c), the phase angle $\phi=0.75 \pi$ and the other parameters are the same as in Fig.~\ref{fig2}(a). The parameters of (d) are those as in Fig.~\ref{fig3}(c) with $g_{\rm ad}/2\pi = 20$ GHz (8 GHz) for the main panel (the inset).}
	\label{fig4}
\end{figure*}

\begin{figure*}[h]
	\centering
	\caption{\textbf{Effect of the excitonic nonlinearity.} Squeezing spectrum (in dB) of the output field quadrature for different values of the Kerr-induced exciton frequency shift $\delta \omega_d/2\pi =$ 0, 1, 2, and 5 GHz (from left to right). We take $G_{\rm db}/2\pi = 4$~GHz and the phase angle $\phi$ is optimized for the squeezing. The other parameters are identical to those in Fig.~\ref{fig2}(a).}
	\label{fig5}
\end{figure*}

\clearpage
\begin{figure*}[p]
	\centering
	\includegraphics[width=0.95\linewidth]{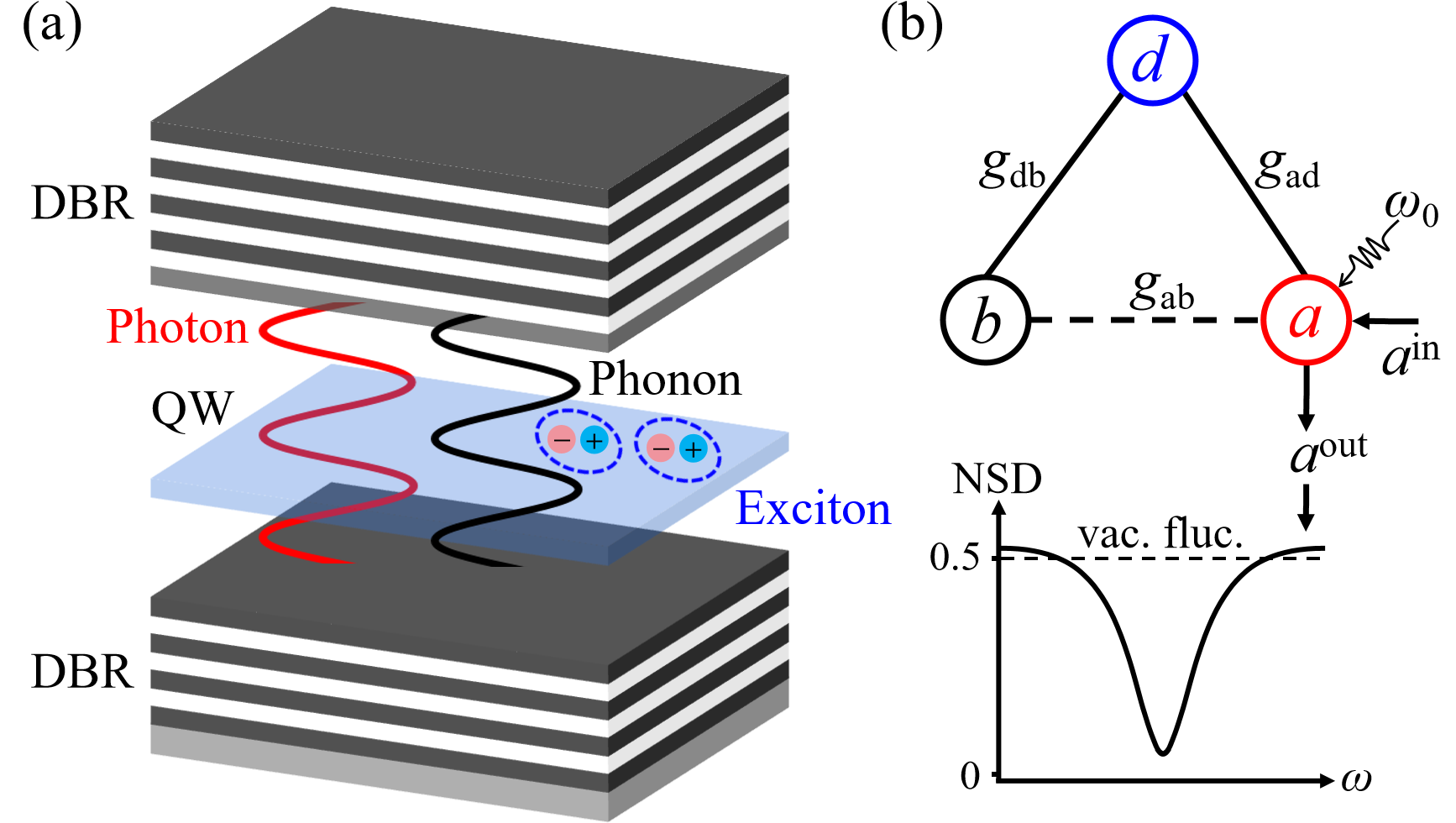}
\end{figure*}

\clearpage
\begin{figure*}[p]
	\centering
	\includegraphics[width=0.95\linewidth]{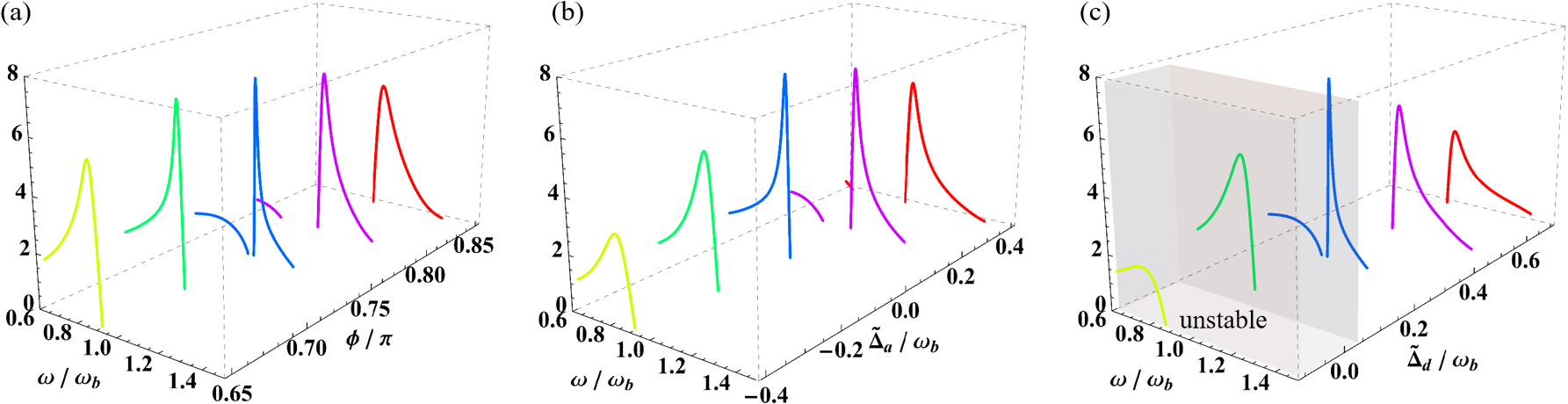}
\end{figure*}

\clearpage
\begin{figure*}[p]
	\centering
	\includegraphics[width=0.9\linewidth]{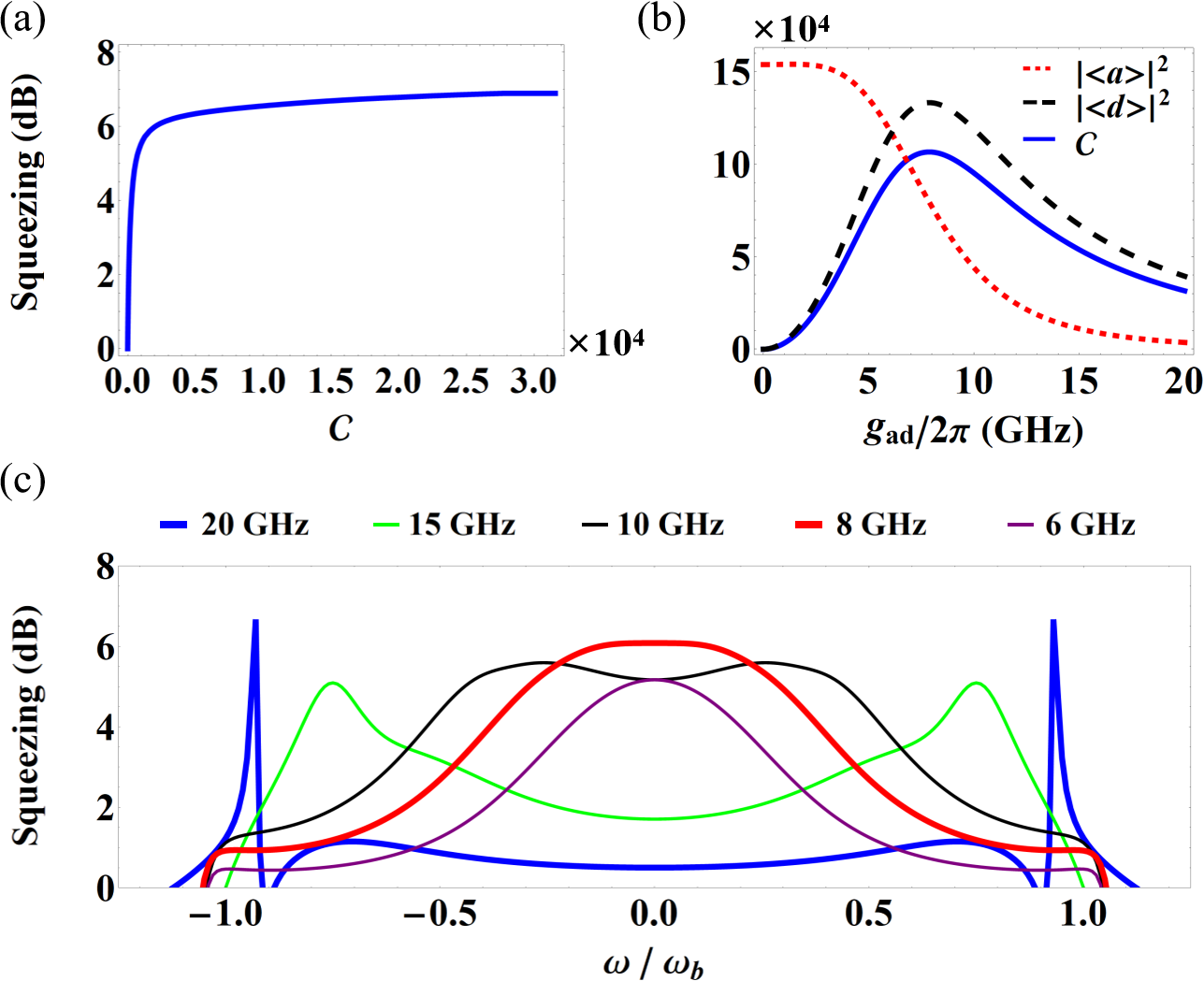}
\end{figure*}

\clearpage
\begin{figure*}[p]
	\centering
	\includegraphics[width=0.9\linewidth]{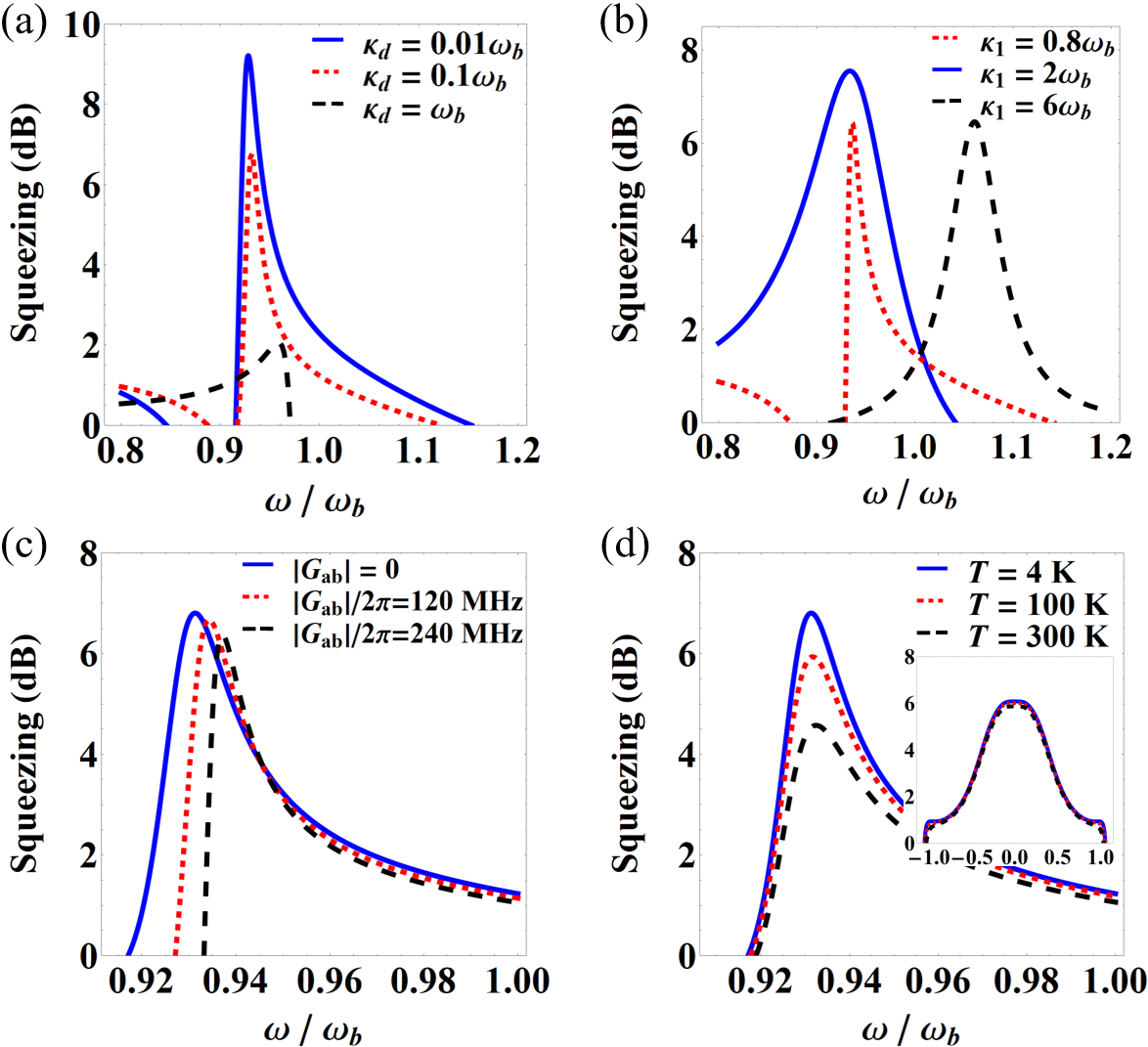}
\end{figure*}

\clearpage
\begin{figure*}[p]
	\centering
	\includegraphics[width=0.96\linewidth]{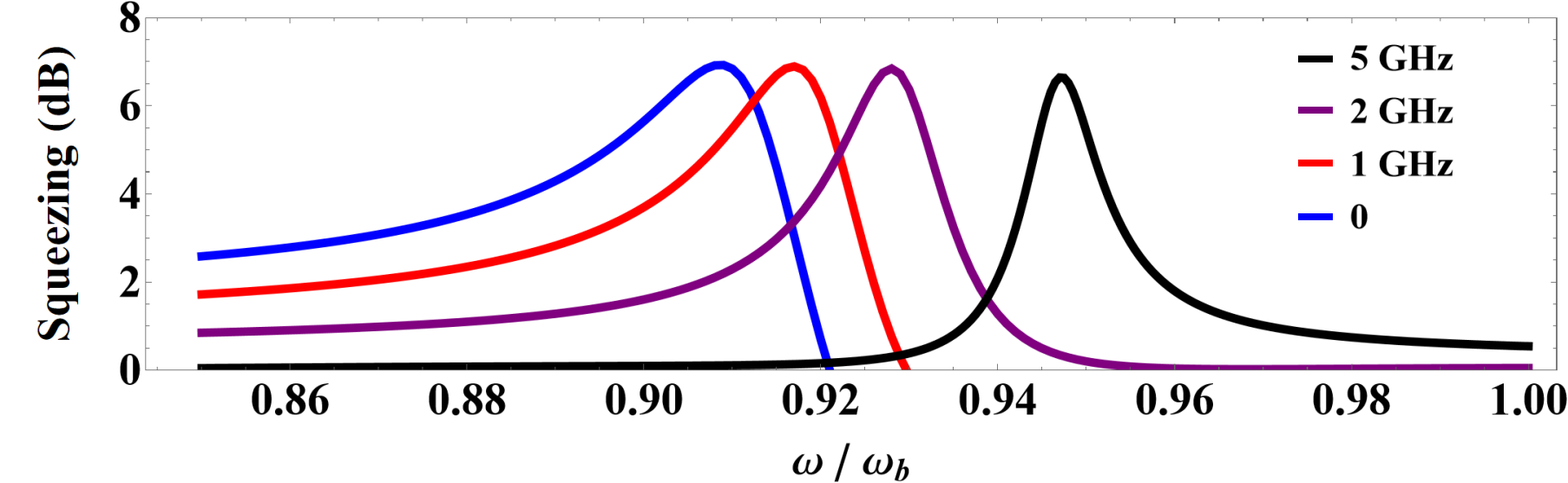}
\end{figure*}

\end{document}